\documentclass[journal,draft,onecolumn,10pt]{IEEEtran}
\usepackage{latexsym,,amssymb,amsmath,graphicx,epsf,cite,bbm,float}
\IEEEoverridecommandlockouts

\def\ninept{\def\baselinestretch{1.58}}
\ninept

\newcommand{\be}{\begin{equation}}
\newcommand{\ee}{\end{equation}}
\newcommand{\bean}{\begin{align}}
\newcommand{\eean}{\end{align}}
\newcommand{\nn}{\nonumber}
\newcommand{\tr}{\mathrm{Tr}}
\newcommand{\veco}{\mathrm{vec}}
\newcommand{\diag}{\mathrm{diag}}

\providecommand{\norm}[1]{\lVert#1\rVert}

\renewcommand{\vec}[1]{ \mbox{$\mathbf {#1}$}}
\renewcommand{\rho}{\mbox{$\delta$}}
\renewcommand{\lambda}{\mbox{$\gamma$}}
\renewcommand{\eta}{\mbox{$\xi$}}

\newcommand{\tA}{\mbox{$\tilde{\vec{H}}$}}
\newcommand{\ty}{\mbox{$\tilde{\vec{y}}$}}
\newcommand{\mA}{\mbox{$\vec{H}$}}
\newcommand{\mB}{\mbox{$\vec{B}$}}
\newcommand{\mQ}{\mbox{$\vec{Q}$}}

\newcommand{\va}{\mbox{\boldmath$\alpha$}}
\newcommand{\vc}{\vec{c}}
\newcommand{\vbet}{\mbox{\boldmath$\beta$}}
\newcommand{\vy}{\mbox{$\vec{y}$}}
\newcommand{\vz}{\mbox{$\vec{z}$}}
\newcommand{\ve}{\mbox{$\vec{e}$}}
\newcommand{\vx}{\mbox{$\vec{u}$}}
\newcommand{\vu}{\mbox{$\vec{u}$}}
\newcommand{\vv}{\mbox{$\vec{v}$}}
\newcommand{\vw}{\mbox{$\vec{w}$}}

\newcommand{\fr}{{\cal R}}

\newcommand{\dy }{\Delta\vec{y}}
\newcommand{\vda}{\vec{h}}
\newcommand{\vb }{\vec{b}}
\newcommand{\vh }{\vec{h}}
\newcommand{\mI }{\vec{I}}
\newcommand{\mM }{\vec{G}}
\newcommand{\mV }{\vec{V}}
\newcommand{\mD }{\vec{\Sigma}}
\newcommand{\mJ }{\vec{D}}
\newcommand{\mX }{\vec{U}}
\newcommand{\mU }{\vec{U}}

\newcommand{\mW }{\vec{W}}
\newcommand{\dA }{\Delta\vec{H}}
\newcommand{\mP }{\vec{P}^{\perp}}
\newcommand{\mPa }{\vec{P}^{\perp}_{\vec{\alpha}}}
\newcommand{\MB}{\left[\begin{array}}
\newcommand{\ME}{\end{array}\right]}

\newcommand{\defi}{\stackrel{\bigtriangleup}{=}}

\begin{document}

\title{A Novel Robust Approach to Least Squares Problems with Bounded Data Uncertainties}
\vspace{-0.5cm}\author{Nargiz Kalantarova, Mehmet A. Donmez, and
  Suleyman S. Kozat, \emph{Senior Member}, IEEE \thanks{Suleyman
    S. Kozat, Nargiz Kalantarova, and Mehmet A. Donmez ({skozat,
      nkalantarova, medonmez}@ku.edu.tr) are with the EE Department at
    the Koc University, Istanbul, tel:+902123381000.}}  \maketitle
\begin{abstract}
In this correspondence, we introduce a minimax regret criteria to the
least squares problems with bounded data uncertainties and solve it
using semi-definite programming. We investigate a robust minimax least
squares approach that minimizes a worst case difference regret. The
regret is defined as the difference between a squared data error and
the smallest attainable squared data error of a least squares
estimator. We then propose a robust regularized least squares approach
to the regularized least squares problem under data uncertainties by
using a similar framework. We show that both unstructured and
structured robust least squares problems and robust regularized least
squares problem can be put in certain semi-definite programming
forms. Through several simulations, we demonstrate the merits of the
proposed algorithms with respect to the the well-known alternatives in
the literature.
\end{abstract}
\begin{center}
\bfseries EDICS Category: SPC-DETC
\end{center}
\section{Introduction}
We study estimation of a deterministic signal observed through an
unknown deterministic data matrix under additive noise
\cite{Ghaoui97}. Although the observation matrix and the output vector
are not exactly known, estimates for both the data matrix and the
output vector, as well as uncertainty bounds on them, are given
\cite{pilanci10, sayed02}.  When there are uncertainties in the model
parameters, a common approach to estimate the input signal is to use
the robust least squares (LS) method \cite{Ghaoui97}, since the classical
LS estimators perform poorly when the perturbations on the model
parameters are relatively high \cite{Ghaoui97, sayed02,
  pilanci10}. Although the robust LS methods are able to minimize the
data error for the worst case perturbations, however, they usually
provide unsatisfactory results on the average
\cite{pilanci10}. Therefore, in order to counterbalance the
conservative nature of the robust LS methods \cite{Ghaoui97}, we
introduce a novel robust LS approach that minimizes a worst case
``regret" that is defined as the difference between the squared data
error and the smallest attainable squared data error with an LS
estimator. By this regret formulation, we seek a linear estimator
whose performance is as close as possible to that of the optimal
estimator for all possible perturbations on the data matrix and the
output vector. Our main goal in proposing the minimax regret
formulation is to provide a trade-off between the robust LS methods
tuned to the worst possible model parameters (under the uncertainty
bounds) and the optimal LS estimator tuned to the underlying unknown
model parameters.  Furthermore, after studying the data estimation
problems in the presence of bounded data uncertainties, we extend the
regret formulation to the regularized LS problem, where the regret is
defined as the difference between the cost of the regularized LS
algorithm \cite{Kailath:book, sayed02}, and the smallest attainable
cost with a linear regularized LS estimator. Under these
frameworks, we provide the solutions for the proposed regret based minimax LS and
the regret based minimax regularized LS approaches in
semi-definite programming (SDP) forms. We emphasize that SDP problems
can be efficiently solved even for real-time applications \cite{boyd}.

A wide range of applications in the signal processing literature deal
with LS problems \cite{sayed02, pilanci10, arikan11, Ghaoui97}.
However, in different applications, the performance of the LS
estimators may substantially degrade due to possible errors in the
model parameters. One of the conventional approaches to find robust
solutions to such estimation problems is the robust LS method
\cite{Ghaoui97, sayed02}, in which the uncertainties in the data
matrix and the output vector are incorporated in optimization via a
minimax residual formulation. Another approach to compensate for data
errors is the total least squares method \cite{pilanci10}, which may
be undesirable since it may yield conservative results due to data
de-regularization. Furthermore, in many linear regression problems,
the data matrix has a known special structure, e.g., Toeplitz
\cite{Ghaoui97, pilanci10}. The performance of the minimax methods
usually improve when such a prior knowledge is integrated in the problem
formulation \cite{Ghaoui97, pilanci10}.

In order to mitigate the pessimistic nature of the worst case
optimization methods, the minimax regret approaches have been
introduced in the context of statistical signal processing literature
\cite{ElMe04, ElTaNe04, eldar_mrr, Kozat:09TSP}. However, we emphasize
that the robust methods studied in this correspondence substantially
differ from \cite{Ghaoui97, sayed02, ElMe04, ElTaNe04,
  Kozat:09TSP}. The cost functions studied here are different than
\cite{Ghaoui97, sayed02}, where the regret terms are directly appended
in the cost functions.  Although in \cite{ElMe04, ElTaNe04,
  Kozat:09TSP} a similar regret notion is used, the cost function as
well as the constraints are substantially different in this
correspondence. Furthermore, note that in \cite{ElMe04}, the
uncertainty is in the statistics of the transmitted signal. On the
other hand, in \cite{ElTaNe04} and \cite{Kozat:09TSP}, the uncertainty
is in the transmitted signal and the channel parameters, respectively,
unlike in here. In this correspondence, the uncertainty is both on the
data matrix and the output vector.  Furthermore, since the cost
functions are different for our formulations, the solutions to the LS
problems presented in this correspondence cannot be obtained from
\cite{ElMe04, ElTaNe04, Kozat:09TSP}. We emphasize that the proposed
methods are formulated for given perturbation bounds and note that
such bounds heavily depend on the estimation algorithms used to obtain
the observation matrix and the data matrix. In this sense, different
estimation algorithms can be readily incorporated in our framework
with the corresponding uncertainty bounds \cite{sayed02}.

Our main contributions are as follows. We first introduce a novel LS
approach in which we seek to find the transmitted data by minimizing
the worst case difference between the data error of the LS estimator
and the data error with the optimal LS estimator. With this method, we
aim to provide a trade off between the performance of the robust LS
methods and the tuned LS estimator (tuned to the unknown data matrix
and the output vector). Then, we develop a minimax regret approach for
the regularized LS problem. We demonstrate that the proposed robust
methods can be cast as SDP problems. In our simulations, we observe
that the proposed approaches yield better performance compared to the
robust methods optimized with respect to the worst case data error
\cite{Ghaoui97, sayed02}, and the tuned LS and the tuned regularized
LS estimators (tuned to the estimated data matrix and the output
vector), respectively.

The organization of the correspondence is as follows. We initially
provide an overview of the problem in Section \ref{sec:system}. In
Section \ref{sec:rls}, we introduce the unstructured and regularized
LS approaches based on the regret formulation. We then continue to
study the structured LS approach. The numerical examples are given in
Section \ref{sec:numer}. Finally, the correspondence concludes with
some remarks in Section \ref{sec:conc}.
\section{System Overview \label{sec:system}}
We consider the problem of estimating a deterministic vector $\vx \in
\mathbb{R}^{n}$ which is observed through a deterministic data
matrix. However, instead of the actual coefficient matrix and the
output vector, their estimates $\mA \in \mathbb{R}^{m\times n}$ and
$\vy \in \mathbb{R}^{m}$ and uncertainty bounds on the estimates are
provided, where $m\geq n$\footnote{All vectors are column vectors and
  denoted by boldface lowercase letters. Matrices are represented by
  boldface uppercase letters. For a vector $\vx$, $\|\vx\| =
  \sqrt{\vx^T \vx}$ is the $\ell_2$-norm, where $\vx^T$ is the
  ordinary transpose. For a matrix $\mA$, $\norm{\mA}$ implies the
  Frobenius norm. For a vector $\vx$, $\mathrm{diag}(\vx)$ is a
  diagonal matrix constructed from the entries of $\vx$. For a square
  matrix $\vec{H}$, $\mathrm{Tr}(\vec{H})$ is the trace. \vec{0}
  denotes a vector or matrix with all zero elements and the dimension
  can be understood from the context. The operator
  $\mathrm{vec}(\cdot)$ stacks the columns of a matrix of dimension
  $m\times n$ into a $mn \times 1$ column vector, and the operator
  $\otimes$ is the Kronecker product \cite{Graham_matcal}.}. Our goal
is to find a solution to the data estimation problem $\mA\vx \approx
\vy$, such that $(\mA+\dA)\vx = \vy+\dy$ for deterministic
perturbations $\dA \in \mathbb{R}^{m\times n}$, $\dy \in
\mathbb{R}^{m}$, which are unknown but a bound on each of the
perturbation is provided, i.e., $\norm{\dA}\leq \rho_H$ and
$\norm{\dy}\leq \rho_Y$, where $\rho_H, \rho_Y \geq 0$.

Even under the uncertainties, the unknown vector $\vx$ can be naively
estimated by simply substituting the estimates $\mA$ and $\vy$ into
the LS estimator \cite{Kailath:book}. For the LS estimator we have
$\hat{\vx}=\mA^+\vy$, where $\mA^+$ is the pseudo-inverse of $\mA$
\cite{Graham_matcal}. However, this approach does not yield acceptable
results, when the errors in the estimates are relatively high
\cite{Ghaoui97, ElMe04, ElTaNe04}. A common approach to find a robust
solution is to use a worst case residual optimization \cite{Ghaoui97}
as
\begin{equation}
\hat{\vx} = \mathrm{arg }\min_{\vx} \max_{\norm{\dA} \leq \rho_H, \norm{\dy} \leq \rho_Y,}{\norm{(\mA + \dA)\vx-(\vy+\dy)}}^2,\nn
\end{equation}
where $\vx$ is found by minimizing the worst case error in the
uncertainty region. However, the solution may be highly conservative,
since the solution is found with respect to the worst possible
coefficient matrix and output vector in the uncertainty regions
\cite{pilanci10, ElMe04, ElTaNe04}. In order to alleviate the
conservative nature of the worst case residual approach as well as to
preserve robustness, we propose a novel LS approach, which provides a
trade off between performance and robustness \cite{ElMe04,
  ElTaNe04}. The regret for not using the optimal LS estimator is
defined as the difference between the squared data error with an
estimate of the input vector and the squared data error with the
optimal LS estimator \be \fr(\dA,\dy) \defi {\norm{(\mA +
    \dA)\vx-(\vy+\dy)}}^2-\min_{\vw}{\norm{(\mA
    +\dA)\vw-(\vy+\dy)}}^2, \nn 
\ee 
where
$\norm{\dA}\leq \rho_H$, $\norm{\dy}\leq \rho_Y$, for some $\rho_H,
\rho_Y \geq 0$.  In the next section, the proposed approaches to the
LS problems are provided. First, the regret based unstructured LS
method is introduced. Then, the unstructured regularized LS approach
is presented in which the worst case regret is optimized, where the
regret is defined as the difference between the cost function of the
regularized LS algorithm \cite{Kailath:book} with an input vector and
the cost function with the optimal regularized LS estimator. Finally,
the structured LS approach is studied.
\section{Robust Least Squares Methods}
\label{sec:rls}
\subsection{Unstructured Robust Least Squares \label{sec:urls}}
In this section, we develop a novel unstructured LS estimator based on
a certain minimax criteria. Let $\mA \in \mathbb{R}^{m\times n}$, $\vy
\in \mathbb{R}^{m}$, $m\geq n$, and $\rho_H, \rho_Y \geq 0$. We find
$\vx$ that is the solution to the following optimization problem:
\begin{equation}
\min_{\vx} \max_{\norm{\dA} \leq \rho_H,\norm{\dy} \leq \rho_Y} \big\{
    {\norm{(\mA + \dA)\vx-(\vy+\dy)}}^2-\min_{\vw}{\norm{(\mA
        +\dA)\vw-(\vy+\dy)}}^2 \big\}. \label{prudef1}
\end{equation}
Note that by unconstraint minimization over $\vw$, we have $g(\tA,
\ty) \defi \norm{(\mI-\tA\tA^{+})\ty}^2=\min_{\vw}{\norm{(\mA
    +\dA)\vw-(\vy+\dy)}}^2$, $\tA \defi \mA + \dA$, $\ty \defi \vy+\dy$. If we
use the first order Taylor series expansion \cite{Graham_matcal}
for $g(\tA, \ty)$, then
\begin{align}
g(\tA, \ty)=g(\mA,\vy) + 2\mathrm{Tr}\bigg\{ \nabla g(\tA, \ty)|_{\tA=\mA, \ty = \vy}^T\big[\dA \hspace{0.1in}\dy\big] \bigg\} + O\big(\norm{\big[\dA \hspace{0.1in} \dy \big]}^2\big).
\label{1stTaylorApp}
\end{align}
Clearly, the effect of this approximation vanishes as
$\norm{\big[\dA \hspace{0.1in} \dy \big]}$ decreases, however, we 
observe through our simulations that even for relatively large
perturbations a satisfactory performance is obtained. Since
$\dfrac{\partial g(\tA, \ty)}{\partial \ty}\vert_{\tA=\mA,
  \ty=\vy}=2(\mI-\mA\mA^+)\vy$ and based on the approximation
\eqref{1stTaylorApp}, the regret in \eqref{prudef1} can be written as
\begin{align}
\fr(\dA, \dy) &\approx {\norm{(\mA + \dA)\vx-(\vy+\dy)}}^2 - g(\mA,\vy) - 2\mathrm{Tr}\bigg\{ \nabla g(\tA, \ty)|_{\tA=\mA, \ty = \vy}^T\big[\dA\hspace{0.1in}\dy \big] \bigg\} \nn\\
&= {\norm{(\mA + \dA)\vx-(\vy+\dy)}}^2-\eta-2\big(\veco (\mJ)^T\veco (\dA)+2\dy^T\mP_{\mA}\vy+2\vy^T\mP_{\mA}\dy\big) \nn\\
&= {\norm{(\mA + \dA)\vx-(\vy+\dy)}}^2-\big(\eta + \vc^T\vda+\vda^T\vc+\dy^T\vb+\vb^T\dy\big), \label{thmReg}
\end{align}
where $\mJ \defi \dfrac{\partial g(\tA, \ty)}{\partial \tA}\vert_{\tA=\mA, \ty=\vy}$, $\mP_{\mA} \defi (\mI-\mA\mA^{+})$ is the projection matrix of the
space perpendicular to the range space of $\mA$, $\eta \defi f(\mA,\vy)$, $\vda \defi \veco (\dA)$, $\vc \defi \veco (\mJ)$ and $\vb \defi 2\mP_{\mA}\vy$.

In the following theorem, we illustrate how the problem of minimization of the worst case regret \eqref{thmReg} can
be put in an SDP form.\\
\noindent{\bf Theorem 1}:
Let $\mA \in \mathbb{R}^{m\times n}$, $\vy \in \mathbb{R}^{m}$, $m\geq n$, and $\rho_H, \rho_Y \geq 0$, then
\be
\min_{\vx} \max_{\norm{\dA} \leq \rho_H,\norm{\dy} \leq \rho_Y} \bigg\{
{\norm{(\mA + \dA)\vx-(\vy+\dy)}}^2 - f(\mA,\vy) - 2\mathrm{Tr}\big\{ \nabla g(\tA, \ty)\vert_{\tA=\mA, \ty=\vy}^T\big[\dA \hspace{0.1in} \dy \big] \big\}
\bigg\} \label{prudefthm1}
\ee
is equivalent to solving the following SDP problem
\be
\min \lambda \mbox{ subject to}\nn
\ee
\be
\left[ \begin{array}{cccc}
 \lambda+ \eta -\tau - \theta & \left(\mA\vx-\vy\right)^T & \rho_H \vc^T & \rho_Y \vb^T\\
 \left(\mA\vx-\vy\right)  & \mI & \rho_H \mX & -\rho_Y\mI \\
 \rho_H\vc & \rho_H\mX^T & \tau\mI & \vec{0} \\
 \rho_Y \vb & -\rho_Y\mI & \vec{0} & \theta \mI
 \end{array}  \right]
  \geq 0, \label{sdpesol1}
\ee
where $g(\tA, \ty)=\ty^T(\mI-\tA\tA^+)\ty$, $\eta = \vy^T(\mI-\mA\mA^+)\vy$, $\tau, \theta \geq 0$, $\vc = \veco \left(\dfrac{\partial g(\tA, \ty)}{\partial \tA}\vert_{\tA=\mA, \ty=\vy}\right)$, $\vb = 2\mP_{\mA}\vy$, $\mX$ is an $m\times mn$ matrix and defined as $\mX \defi \mI \otimes \vx^T$.\\
\noindent{\bf Proof}: The proof is in the Appendix. $\square$

As the special case, the pseudo-inverse of $\mA$ is $\mA^+ =
(\mA^T\mA)^{-1}\mA^T$, when $\mA$ is full rank. For this case, we
introduce the next lemma to explicitly calculate $\nabla f(\tA, \ty)$.

\noindent{\bf Lemma 1}: Let $\mA\in\mathbb{R}^{m\times n}$, $m \geq n$, be a full rank matrix, $\vy\in \mathbb{R}^n$ and define
 $g(\mA, \vy) \defi \vy^T(\mI-\mA\mA^{+})\vy$, then
\be
\dfrac{\partial g(\mA, \vy)}{\partial \mA}= -2\left( \mA(\mA^T\mA)^{-1}\mA^T\vy\vy^T\mA(\mA^T\mA)^{-1}+\vy\vy^T\mA(\mA^T\mA)^{-1}\right).
\label{lemma1ures}
\ee
{\bf Proof}: Taking the partial derivative of $f(\mA, \vy)$ with respect to $\mA_{ij}$ based on \cite{Graham_matcal} yields
\begin{align}
\dfrac{\partial g(\mA, \vy)}{\partial \mA_{ij}}&=-\vy^T\left[\ve_i\ve_l^T(\mA^T\mA)^{-1}\mA^T-2\mA(\mA^T\mA)^{-1}\mA^T\ve_i\ve_j^T(\mA^T\mA)^{-1}\mA^T+
\mA(\mA^T\mA)^{-1}\ve_j\ve_i^T\right]\vy \nn\\
&=2\ve_j^T(\mA^T\mA)^{-1}\mA^T\vy\vy^T\ve_i-2\ve_j^T(\mA^T\mA)^{-1}\mA^T\vy\vy^T\mA(\mA^T\mA)^{-1}\mA^T\ve_i \label{Partialy1u}.
\end{align}
By definition, the transpose of the term in \eqref{Partialy1u} is the $ij^\mathrm{th}$ entry of the matrix \\ $-2\left[\vy\vy^T\mA(\mA^T\mA)^{-1} - \mA(\mA^T\mA)^{-1}\mA^T\vy\vy^T\mA(\mA^T\mA)^{-1}\right]$, hence the result in \eqref{lemma1ures} follows. $\square$ \\
From\cite{Graham_matcal}, we have
\be
\dfrac{\partial g(\mA,\vy)}{\partial \vy}= 2(\mI-\mA\mA^{+})\vy = 2 \mP_{\mA} \vy,
\label{derivy}
\ee
where $\mP_{\mA}=(\mI-\mA\mA^{+})$.
By using Lemma 1 and \eqref{derivy} in \eqref{1stTaylorApp}, we get
\begin{align}
g(\tA, \ty)& \approx \eta + 2\mathrm{Tr}\left\{
\begin{bmatrix}-(\vy\vy^T\mA(\mA^T\mA)^{-1} + \mA\mA^+\vy\vy^T\mA(\mA^T\mA)^{-1}) & \mP\vy
\end{bmatrix}^T\begin{bmatrix}\dA & \dy \end{bmatrix} \right\}\nn\\
& = \eta -4 \left[\left( \vy^T\dA \vv + \vy^T \mA\mA^+\dA\vv \right) -\vy^T\mP_{\mA}\dy\right]\nn\\
& = \eta -4 \left[\left( \vy^T\mV \vda + \vy^T \mA^+\mV\vda \right) -\vy^T\mP_{\mA}\dy\right]\label{lineF4}\\
& = \eta + \vz^T\vda + \vda^T\vz + 2\dy^T\mP_{\mA}\vy + 2\vy^T\mP_{\mA}\dy, \nn 
\end{align}
where $\eta = \vy^T(\mI-\mA\mA^{+})\vy$, $\vda = \mathrm{vec} \left({\dA^T}\right)$, $\vv =\mA^+\vy $, and $\vz = -2(\vy^T\mV + \vy^T
\mA(\mA^T\mA)^{-1}\mA^T\mV)^T$. Equation \eqref{lineF4} follows since $\dA\vv=\mV\vda$ and $\mB$ is an $m\times mn$ matrix defined as $\mV \defi \mI \otimes \vv^T$. 

If $\mA$ is not full rank, then the pseudo-inverse of $\mA$ can be written
as $\mA^+ = \mV\mD^+\mU^T$ by using the singular value decomposition
\cite{Graham_matcal}, where $\mU$ is an $m \times n$ unitary matrix,
$\mD$ is an $n\times n$ diagonal matrix with nonnegative real numbers
on the diagonal, and $\mV$ is an $n\times n$ unitary matrix. The
nonzero diagonal entries of $\mD \defi \diag
[\sigma_1,\sigma_2,\ldots,\sigma_r,0,\ldots,0]$ are known as the
singular values of $\mA$. In order to calculate $\nabla g(\tA,
\ty)\vert_{\tA=\mA, \ty=\vy}$, it is sufficient to calculate
$\dfrac{\partial \mA^+}{\partial A_{ij}}$. Then, the derivations
follow the full rank case using \be \dfrac{\partial \mA^+}{\partial
  H_{ij}}=\dfrac{\partial \mV}{\partial
  H_{ij}}\mD^+\mU^T+\mV\dfrac{\partial \mD^+}{\partial
  H_{ij}}\mU^T+\mV\mD^+\dfrac{\partial \mU}{\partial H_{ij}}^T =
\dfrac{\partial \mV}{\partial
  H_{ij}}\mD^+\mU^T+\mV\mD^{-1}\dfrac{\partial \mD}{\partial
  H_{ij}}\mD^{-1}\mU^T +\mV\mD^+\dfrac{\partial \mU}{\partial
  H_{ij}}^T,\nn \ee where the partial derivatives of $\mU$, $\mV$, and
$\mD$ with respect to $ij^{\mathrm{th}}$ element of $\mA$ are derived
in \cite{Papado00}.
\subsection{Unstructured Robust Regularized Least Squares}
\label{sec:crls}
A wide range of applications in signal processing literature require
solutions to regularized LS problems \cite{Kailath:book}. In
\cite{sayed02}, a worst case optimization approach is developed to
solve the regularized LS problem. To reduce the conservative nature of
\cite{sayed02}, we next develop a regret based regularized LS approach
when the model parameters are subject to uncertainties. The regret for
not using the optimal regularized LS method is defined as the
difference between the cost function of the regularized LS algorithm
with an estimate of the input vector and the cost function of the
regularized LS algorithm with the regularized LS estimator as \be
\fr(\dA,\dy) \defi {\norm{(\mA +
    \dA)\vx-(\vy+\dy)}}^2+\mu\norm{\vx}^2-\min_{\vw}\{\norm{(\mA
  +\dA)\vw-(\vy+\dy)}^2+\mu\norm{\vw}^2\}, \nn \label{RLSreg0} \ee
where $\norm{\dA}\leq \rho_H$, and $\norm{\dy}\leq \rho_Y$ for some
$\rho_H, \rho_Y \geq 0$, and $\mu > 0$ is a regularization
parameter. We emphasize that there are different approaches to choose
$\mu$, however, for the focus of this correspondence, we assume that
it is already set before the optimization. Hence, we solve the
the regularized LS problem for any given value of $\mu > 0$.

Given $\mA \in \mathbb{R}^{m\times n}$ and $m\geq n$, $\vy \in \mathbb{R}^{m}$ we have
\be
\min_{\vx} \max_{\norm{\dA} \leq \rho_H,\norm{\dy} \leq \rho_Y} \bigg\{
{\norm{(\mA + \dA)\vx-(\vy+\dy)}}^2+ \mu \norm{\vx}^2
-\min_{\vw}\big\{{\norm{(\mA +\dA)\vw-(\vy+\dy)}}^2+\mu \norm{\vw}^2\big\}
\bigg\}. \label{prRLSdef}
\ee
Since $\tA = \mA + \dA$, $\ty = \vy + \dy$ and inserting the optimal regularized LS solution in \eqref{prRLSdef} yields
\be
\min_{\vx} \max_{\norm{\dA} \leq \rho_H,\norm{\dy} \leq \rho_Y}
\big\{ {\norm{\tA\vx-\ty}}^2+\mu\norm{\vx}^2-{\ty^T(\mI + \mu^{-1}\tA\tA^{T})^{-1}\ty} \big\}.\nn 
\ee
Note that $(\mI + \mu^{-1}\tA\tA^{T})>0$, hence it is invertible.
Denoting $h(\tA, \ty) \defi \ty^T(\mI + \mu^{-1}\tA\tA^{T})^{-1}\ty$, and using the first order Taylor
series expansion \cite{Graham_matcal} for $g(\tA, \ty)$ yields
\begin{align}
h(\tA, \ty)=h(\mA,\vy) + 2\mathrm{Tr}\bigg\{ \nabla h(\tA, \ty)\vert_{\tA=\mA, \ty = \vy}^T\big[\dA \hspace{0.1in} \dy\big] \bigg\} + O\bigg(\norm{\big[\dA \hspace{0.1in} \dy\big]}^2\bigg).
\label{1stTaylorRLS}
\end{align}
The following lemma is introduced to calculate $\nabla h(\tA, \ty)\vert_{\tA=\mA, \ty = \vy}$ in \eqref{1stTaylorRLS}. \\
\noindent{\bf Lemma 2}: Let $\mA\in\mathbb{R}^{m\times n}$, $\vy\in \mathbb{R}^n$ and define
 $h(\mA, \vy) \defi \vy^T(\mI + \mu^{-1}\mA\mA^{T})^{-1}\vy$, then
\be
\dfrac{\partial h(\mA, \vy)}{\partial \mA}= -2\mu^{-1} (\mI + \mu^{-1}\mA\mA^T)^{-1}\vy\vy^T(\mI + \mu^{-1}\mA\mA^T)^{-1}\mA.
\label{lemma3result}
\ee
{\bf Proof}: Taking the
partial derivative of $h(\mA, \vy)$ with respect to $\mA_{ij}$ based on \cite{Graham_matcal} yields
\begin{align}
\dfrac{\partial h(\mA, \vy)}{\partial \mA_{ij}}&=-\vy^T(\mI + \mu^{-1}\mA\mA^T)^{-1}(\mu^{-1}\ve_i\ve_j^T\mA^T+\mu^{-1}\mA\ve_j\ve_i^T)(\mI + \mu^{-1}\mA\mA^T)^{-1}\vy \nn\\
&=-2\mu\ve_i^T(\mI + \mu^{-1}\mA\mA^T)^{-1}\vy\vy^T(\mI + \mu^{-1}\mA\mA^T)^{-1}\mA\ve_j. \label{eq:deriv}
\end{align}
By definition, the term in \eqref{eq:deriv} is the $ij^{\mathrm{th}}$ entry of the matrix $-2\mu(\mI + \mu\mA\mA^{T})^{-1}\vy\vy^T(\mI + \mu\mA\mA^{T})^{-1}\mA$, hence the result \eqref{lemma3result} follows. $\square$\\
Also, from \cite{Graham_matcal} we have
$\dfrac{\partial h(\mA, \vy)}{\partial \vy} = 2(\mI + \mu^{-1}\mA\mA^{T})^{-1}\vy$.
Then, one can write the regret in \eqref{RLSreg0} based on the expansion in \eqref{1stTaylorRLS} as
\be
\fr(\dA, \dy) \approx (\tA\vx-\ty)^T(\tA\vx-\ty)+\mu\vx^T\vx-(\eta+\vc^T\vda+\vda^T\vc+\vb^T\dy+\dy^T\vb), \label{reg_rls}
\ee
where $\eta = \vy^T(\mI+ \mu^{-1}\mA\mA^T)^{-1}\vy$, $\vda = \veco (\dA)$, $\vb = 2(\mI + \mu^{-1}\mA\mA^{T})^{-1}\vy$ and $\vc = \veco \bigg(\dfrac{\partial h(\mA,\vy)}{\partial \mA}\bigg)$.
In the next theorem, we show that the minimization of the worst case regret \eqref{reg_rls} can be put into an SDP form.\\
{\bf Theorem 2}: Let $\mA \in \mathbb{R}^{m\times n}$, $m\geq n$, $\vy \in \mathbb{R}^{m}$, and $\rho_H, \rho_{Y} \geq 0$, then
\be
\min_{\vx} \max_{\norm{\dA} \leq \rho_H,\norm{\dy} \leq \rho_Y} \big\{
{\norm{(\mA + \dA)\vx-(\vy+\dy)}}^2+ \mu \norm{\vx}^2- (\eta+\vc^T\vda+\vda^T\vc+\vb^T\dy+\dy^T\vb)
\big\} \label{prudefthm2}
\ee
is equivalent to solving the following SDP problem
\be \min \lambda \mbox{ subject to}\nn\ee
\be
\left[ \begin{array}{ccccc}
 \lambda+ \eta -\tau - \theta & (\mA\vx-\vy)^T & \vx^T & \rho_H \vc^T & \rho_Y \vb^T\\
 (\mA\vx-\vy)  & \mI & \vec{0} & \rho_H \mX & -\rho_Y\mI  \\
 \vx & \vec{0} & \mu\mI & \vec{0} & \vec{0} \\
 \rho_H\vc & \rho_H\mX^T & \vec{0} & \tau\mI & \vec{0}\\
 \rho_Y \vb & -\rho_Y\mI & \vec{0} & \vec{0} & \theta \mI
 \end{array}  \right]
  \geq 0, \label{sdpRLSsol1}
\ee
where $\mu>0$ is a regularization parameter, $\tau, \theta \geq 0$, $\eta = \vy^T(\mI+ \mu\mA\mA^T)^{-1}\vy$, $\vda = \veco (\dA)$, $\vb = 2(\mI + \mu^{-1}\mA\mA^{T})^{-1}\vy$, $\vc = \mathrm{vec} \bigg( \dfrac{ \partial h(\mA, \vy)}{ \partial \mA} \bigg)$ for $h(\tA, \ty) = \ty^T(\mI + \mu^{-1}\tA\tA^T)^{-1}\ty$ and $\mX$ is an $m\times mn$ matrix constructed by $\vx$, where $\mX \defi \mI \otimes \vx^T$.\\
\noindent{\bf Proof}: The proof follows similar lines with the proof of
Theorem 1, hence omitted here. $\square$
\subsection{Structured Robust Least Squares}
\label{sec:srls}
In many engineering applications the data matrix has a special
structure, e.g., Toeplitz, Hankel, Vandermonde, as well as the
perturbations on them \cite{Ghaoui97, pilanci10}. Leveraging this
prior knowledge may improve the performance of the regret based
minimax LS approach \cite{Ghaoui97, pilanci10}. Therefore, in this
section, we study a special case of the problem \eqref{prudef1}, where
the associated perturbations for $\mA$ and $\vy$ are structured. The
structure on the perturbations is defined as follows as: $\dA =
\sum_{i=1}^p\alpha_i\mA_i$ and $\dy = \sum_{i=1}^p\beta_i\vy_i$, where
$\mA_i \in \mathbb{R}^{m \times n}$, $\vy_i \in \mathbb{R}^m$, $m \geq
n$, and $p$ are known but $\alpha_i,\beta_i \in \mathbb{R}$,
$i=1,\ldots,p$, are unknown. However, the bounds on the norm of $\va
\defi \begin{bmatrix}\alpha_1 & \alpha_2 & \ldots &
  \alpha_p \end{bmatrix}^T$ and $\vbet \defi \begin{bmatrix}\beta_1 &
  \beta_2 & \ldots & \beta_p \end{bmatrix}^T$ are provided as
$\norm{\va} \leq \rho_H, \norm{\vbet} \leq \rho_B$, where $\rho_H,
\rho_B \geq 0$. We emphasize that this formulation can model a wide
range of structural constraints.  We seek to solve the following
optimization problem \be \min_{\vx} \max_{\norm{\va} \leq \rho_H,
  \norm{\vbet} \leq \rho_B} \left[ {\norm{\mA
      (\va)\vx-\vy(\vbet)}}^2-\min_{\vw}{\norm{\mA
      (\va)\vw-\vy(\vbet)}}^2 \right], \label{prdef1} \ee where
$\mA(\va)= \mA + \sum_{i=1}^p \alpha_i\mA_i$, and $\vy (\vbet)= \vy +
\sum_{i=1}^p \beta_i\vy_i$.

We point out that $\min_{\vw}{\norm{\mA (\va)\vw-\vy(\vbet)}}^2 =
\norm{(\mI-\mA (\va)\mA (\va)^{+})\vy(\vbet)}^2$,
where $\mA (\va)^{+}
\defi (\mA (\va)^T\mA (\va))^{-1}\mA (\va)^T$ is the pseudo-inverse of $\mA (\va)$ and $\mPa
\defi (\mI-\mA(\va)\mA(\va)^{+})$ is the projection matrix of the
space perpendicular to the range space of $\mA(\va)$. Here, $\mA$ and $\mA(\va)$ are assumed to be full rank. We use
the first order Taylor series expansion in order to
express the term $\vy(\vbet)^T(\mI-\mA(\va)\mA(\va)^{+})\vy(\vbet)$ as
\begin{align}
\vy(\vbet)^T(\mI-\mA(\va)\mA(\va)^{+})\vy(\vbet) &= \vy^T(\mI-\mA\mA^{+})\vy \nn + 2\tr\{\nabla_{[\va \hspace{2mm} \vbet]}\norm{\mPa \vy}^2\Vert_{[\va \hspace{2mm} \vbet]=\vec{0}}^T[\va \hspace{2mm} \vbet]\}\nn\\ &+ O\left( \norm{[\va \hspace{2mm} \vbet]}^2\right).  \label{TaylorApp}
\end{align}
If we denote the regret term as $\fr(\va, \vbet) = {\norm{\mA (\va)\vx-\vy(\vbet)}}^2-\norm{(\mI-\mA (\va)\mA
(\va)^{+})\vy(\vbet)}^2$, then the regret can be written based on \eqref{TaylorApp} as
\be
\fr(\va, \vbet) \approx {\norm{\mA (\va)\vx-\vy(\vbet)}}^2-\vy^T(\mI-\mA\mA^{+})\vy
-2\tr\{\nabla_{[\va \hspace{2mm} \vbet]}\norm{\mPa \vy}^2\Vert_{[\va \hspace{2mm} \vbet]=\vec{0}}^T[\va \hspace{2mm} \vbet]\}. \label{reg_str}
\ee
We introduce the following lemma to compute the last term in \eqref{reg_str}.\\
\noindent{\bf Lemma 3}: Let $\mA, \mA_1, \ldots, \mA_p \in \mathbb{R}^{m\times n}$, $\vy \in \mathbb{R}^{m}$, $\mA(\va) = \mA + \sum_{i=1}^p
\alpha_i\mA_i$, $\va = [\alpha_1, \alpha_2, \ldots, \alpha_p]$, where $m \geq n$, and let $\mA$ and $\mA(\va)$ be full rank. If we define $g(\mA(\va), \vy) \defi \vy^T(\mI-\mA(\va)\mA(\va)^{+})\vy$,
 then  $\dfrac{\partial g(\mA(\va), \vy)}{\partial \alpha_i} = -2 \vy^T \mPa \mA_i \mA(\va)^{+}\vy$.\\
{\bf Proof}:
By using the result of Lemma 1,
\begin{align}
&\dfrac{\partial g(\mA(\va), \vy)}{\partial \alpha_i} = \mathrm{Tr} \left\{\dfrac{\partial g(\mA(\va), \vy)}{\partial \mA(\va)}^T \dfrac{\partial \mA(\va)}{\partial \alpha_i}\right\} \nn\\
&=\mathrm{Tr}\bigg\{ \bigg(-2( \mA(\va)(\mA(\va)^T\mA(\va))^{-1}\mA(\va)^T\vy\vy^T\mA(\va)(\mA(\va)^T\mA(\va))^{-1}+\vy\vy^T\mA(\va)(\mA(\va)^T\mA(\va))^{-1}) \bigg)^T\bigg(\dfrac{\partial \mA(\va)}{\partial \alpha_i} \bigg)\bigg\}\nn\\
&=
-2 \vy^T \mPa \mA_i \mA(\va)^{+}\vy,\nn
\end{align}
where $\vec{P}_{\va} \defi \mA(\va)\mA(\va)^{+}$ is the projection matrix of the range space of $\mA(\va)$. $\square$\\
From Lemma 2, it follows that
$
\dfrac{\partial }{\partial \alpha_i} \left[ \vy(\vbet)^T \mPa \vy(\vbet) \right] =
-2 \vy(\vbet)^T \mPa \mA_i \mA (\va)^{+} \vy(\vbet). 
$
Also,
$
\dfrac{\partial }{\partial \beta_i} \left[ \vy(\vbet)^T \mPa \vy(\vbet) \right] = 2\vy(\vbet)^T(\mI-\mA(\va)\mA(\va)^{+})\vy_i = 2 \vy(\vbet)^T\mPa \vy_i. 
$
   If we denote $\eta \defi \vy^T(\mI-\mA\mA^{+})\vy$,
$\vb \defi \nabla_{\va}\norm{\mPa \vy(\vbet)}^2|_{\va=\vec{0}, \vbet=\vec{0}}$,
and $\vc \defi \nabla_{\vbet}\norm{\mPa \vy(\vbet)}^2|_{\va=\vec{0}, \vbet=\vec{0}}$,
then based  on \eqref{TaylorApp} we obtain
\be
\vy(\vbet)^T(\mI-\mA(\va)\mA(\va)^{+})\vy(\vbet) \approx \eta + \vb^T\va+ \va^T\vb + \vbet^T\vc+\vc^T\vbet.
\label{Taylor_bet2}
\ee
By using the result \eqref{Taylor_bet2}, in the next theorem we show that
the problem of estimating $\vx$ by minimizing the worst case regret \eqref{reg_str} can be cast as an SDP problem.\\
\noindent{\bf Theorem 3}:
Let $\mA, \mA_1, \ldots, \mA_p \in \mathbb{R}^{m\times n}$, $\vy, \vy_1,\ldots, \vy_p \in \mathbb{R}^{m}$, $\rho_{A}, \rho_{B} \geq 0$, where $m\geq n$, and let $\mA$ and $\mA(\va)$ be full rank, then
\be
  \min_{\vx} \max_{\norm{\va} \leq \rho_H, \norm{\vbet} \leq \rho_B} \big\{{\norm{\mA (\va)\vx-\vy(\vbet)}}^2-\vy^T(\mI-\mA\mA^{+})\vy
-2\tr\{\nabla_{[\va \hspace{2mm} \vbet]}\norm{\mPa \vy}^2\vert_{[\va \hspace{2mm} \vbet]=\vec{0}}^T[\va \hspace{2mm} \vbet]\}
\big\}, \label{prdefcorr1}
\ee
is equivalent to solving the following SDP problem
\be \min \lambda \mbox{ subject to}\nn\ee
\be
\left[ \begin{array}{cccc}
 \lambda+ \eta -\tau -\theta& \left(\mA\vx-\vy\right)^T & \rho_H \vb^T & \rho_B \vc^T \\
  \mA\vx-\vy  & \mI & \rho\mM & -\rho_B \mQ \\
  \rho_H\vb & \rho\mM^T & \tau\mI & \vec{0}\\
  \rho_B \vc & -\rho_B \mQ^T & \vec{0} & \theta\mI
 \end{array}  \right]
  \geq 0,\label{sdpconstrcorr1}
 \ee
where $\va \defi \begin{bmatrix} \alpha_1 &
\alpha_2 & \ldots & \alpha_p \end{bmatrix}^T$, $\mA (\va)= \mA + \sum_{i=1}^p
\alpha_i\mA_i$, $\vbet \defi \begin{bmatrix} \beta_1 &
\beta_2 & \ldots & \beta_p \end{bmatrix}^T$, $\vy (\vbet)= \vy + \sum_{i=1}^p
\beta_i\vy_i$,
$\vb = \nabla_{\va}\norm{\mPa \vy(\vbet)}^2|_{\va=\vec{0}, \vbet=\vec{0}}$,
$\vc \defi \nabla_{\vbet}\norm{\mPa \vy(\vbet)}^2|_{\va=\vec{0}, \vbet=\vec{0}}$,
$\eta = \vy^T(\mI-\mA\mA^{+})\vy$, $\tau, \theta \geq 0$, $\mM \defi \begin{bmatrix} \mA_1\vx & \mA_2\vx & \ldots &
\mA_p\vx \end{bmatrix}$, and $\mQ = \begin{bmatrix} \vy_1 & \ldots & \vy_p \end{bmatrix}$.\\
\noindent{\bf Proof}: The outline of the proof follows similar lines to the proof of Theorem 1, hence it is omitted here. $\square$

In the following corollary, we provide a reduced version of Theorem 3, where the structure of the perturbations and
the bounds on the perturbations are defined using the same parameters as commonly studied in \cite{Ghaoui97, pilanci10}, i.e., if $\dA = \sum_{i=1}^p \alpha_i\mA_i$, then $\dy =  \sum_{i=1}^p \alpha_i\vy_i$,
where $\mA \in \mathbb{R}^{m\times n}$, $\vy \in \mathbb{R}^m$ and $\alpha_i \in \mathbb{R}$ for all $i=1, \ldots, p$.\\
\noindent{\bf Corollary}:
Let $\mA, \mA_1, \ldots, \mA_p \in \mathbb{R}^{m\times n}$, $\vy, \vy_1,\ldots, \vy_p \in \mathbb{R}^{m}$, $\rho\geq 0$,  where $m\geq n$. Assume $\mA$ and $\mA(\va)$ are full rank, then $\min_{\vx} \max_{\norm{\va} \leq \rho} \big\{
{\norm{\mA (\va)\vx-\vy(\va)}}^2-\vy^T(\mI-\mA\mA^{+})\vy
-2\tr\left\{\nabla_{\va}\norm{\mPa \vy}^2\vert_{\va=\vec{0}}^T\va\right\}
\big\}$
is equivalent to solving the following SDP problem
\be \min \lambda \mbox{ subject to}\nn\ee
\be
\left[
\begin{array}{ccc}
\lambda+ \eta -\tau & \left(\mA\vx-\vy\right)^T & \rho \vb^T \\
\mA\vx-\vy  & \mI & \rho\mM  \\
\rho\vb & \rho\mM^T & \tau\mI
\end{array}
\right]
\geq 0, \nn
\ee
where $\va \defi \begin{bmatrix} \alpha_1 &
\alpha_2 & \ldots & \alpha_p \end{bmatrix}^T$, $\mA (\va)= \mA + \sum_{i=1}^p
\alpha_i\mA_i$, $\vy (\va)= \vy + \sum_{i=1}^p
\alpha_i\vy_i$, $\vb \defi \nabla_{\va}\norm{\mPa \vy(\va)}^2|_{\va=\vec{0}}$,
$\mM \defi \begin{bmatrix} \mA_1\vx-\vy_1 & \mA_2\vx-\vy_2 & \ldots &
\mA_p\vx-\vy_p \end{bmatrix}$, $\eta=\vy^T(\mI-\mA\mA^{+})\vy$, and $\tau \geq 0$.
\section{Simulations}
\label{sec:numer}
We provide numerical examples in different scenarios in order to
illustrate the merits of the proposed algorithms. In the first set of the
experiments, we randomly generate a data matrix of size $m \times n$,
and an output vector of size $m\times 1$, which are normalized to have
unit norm. Then, we randomly generate $200$ random perturbations
$\dA$, $\dy$, where $\norm{\dA}\leq \rho_H$, $\norm{\dy}\leq \rho_Y$,
$m=5$, $n=3$, and $\rho_H=\rho_Y=0.4$. Here, we label the algorithm in
Theorem 1 as ``{\it c-LS}", the robust LS algorithm of \cite{Ghaoui97}
as ``{\it r-LS}", and finally the LS algorithm tuned to the estimates
of the data matrix and the output vector as ``{\it LS}" where we
solve $(\mA+\dA)\vx =\vy + \dy$. In Fig. 1, we plot the corresponding
sorted errors in ascending order. Since the r-LS algorithm optimizes
the worst case squared data error with respect to worst possible
disturbance, it yields the smallest worst case squared error among all
algorithms for these simulations. The largest errors are $1.585$ for
the LS algorithm, $1.267$ for the c-LS algorithm and $1.120$ for the
r-LS algorithm. Nevertheless, the overall performance of the r-LS
algorithm is significantly inferior to the LS and the r-LS algorithms
due to its highly conservative nature. Furthermore, we notice that the
c-LS algorithm provides superior average performance compared to the
LS and the r-LS algorithms, and superior worst case error compared to
the LS algorithm for these simulations.
\begin{figure*}[t!]
\hfill
\begin{minipage}[t]{.46\textwidth}
  \begin{center}
  \vspace{-0.1cm}
  \epsfxsize=3.3in {\epsfbox{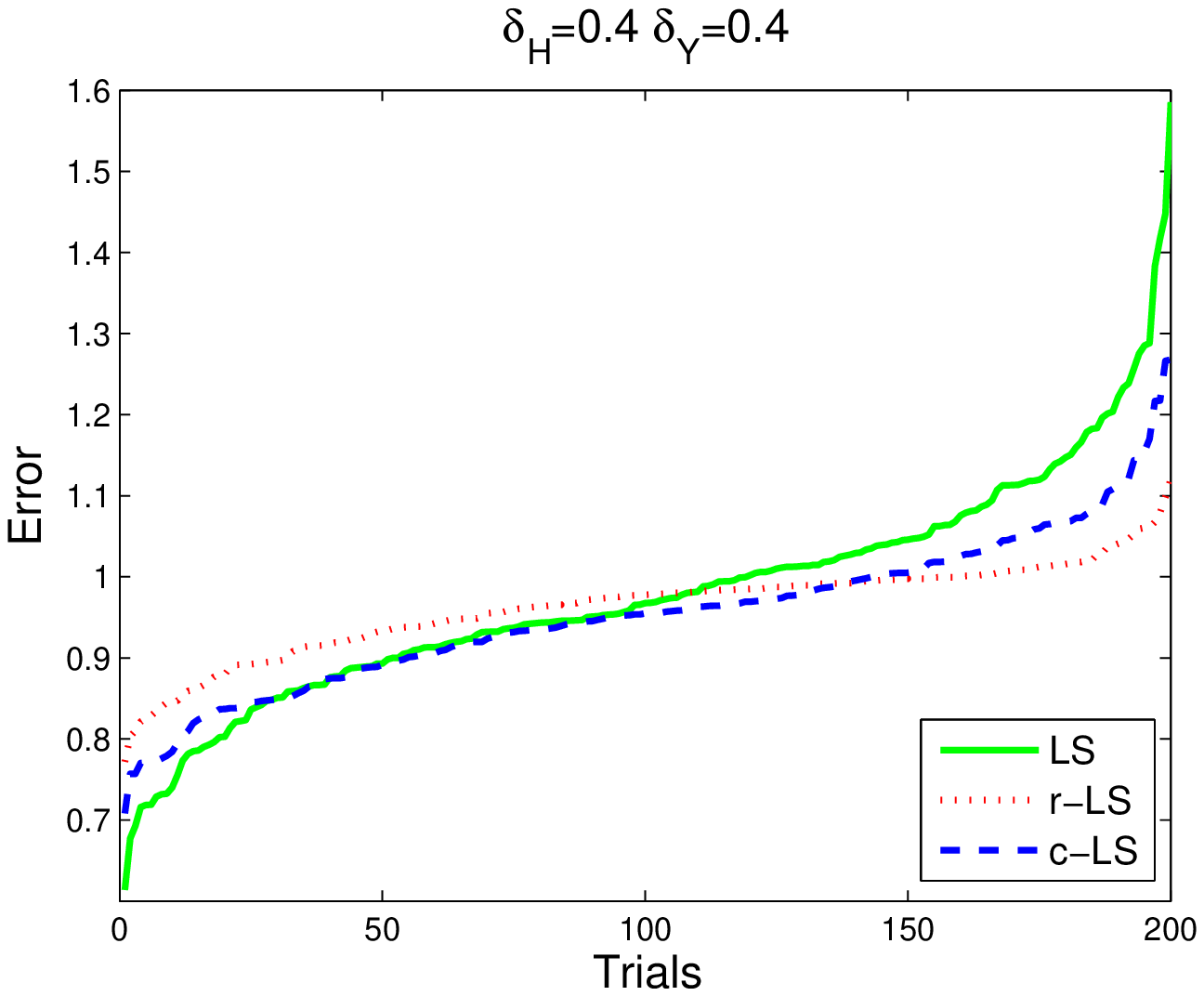}}
  \caption{\small Sorted errors for r-LS, c-LS and LS algorithms over $200$ trials when $\rho_H=\rho_Y=0.4$.}  \vspace{-0.2cm}
  \label{fig:1}
  \end{center}
\end{minipage}
\hfill
\begin{minipage}[t]{.46\textwidth}
  \begin{center}
  \vspace{-0.1cm}
     \epsfxsize=3.3in {\epsfbox{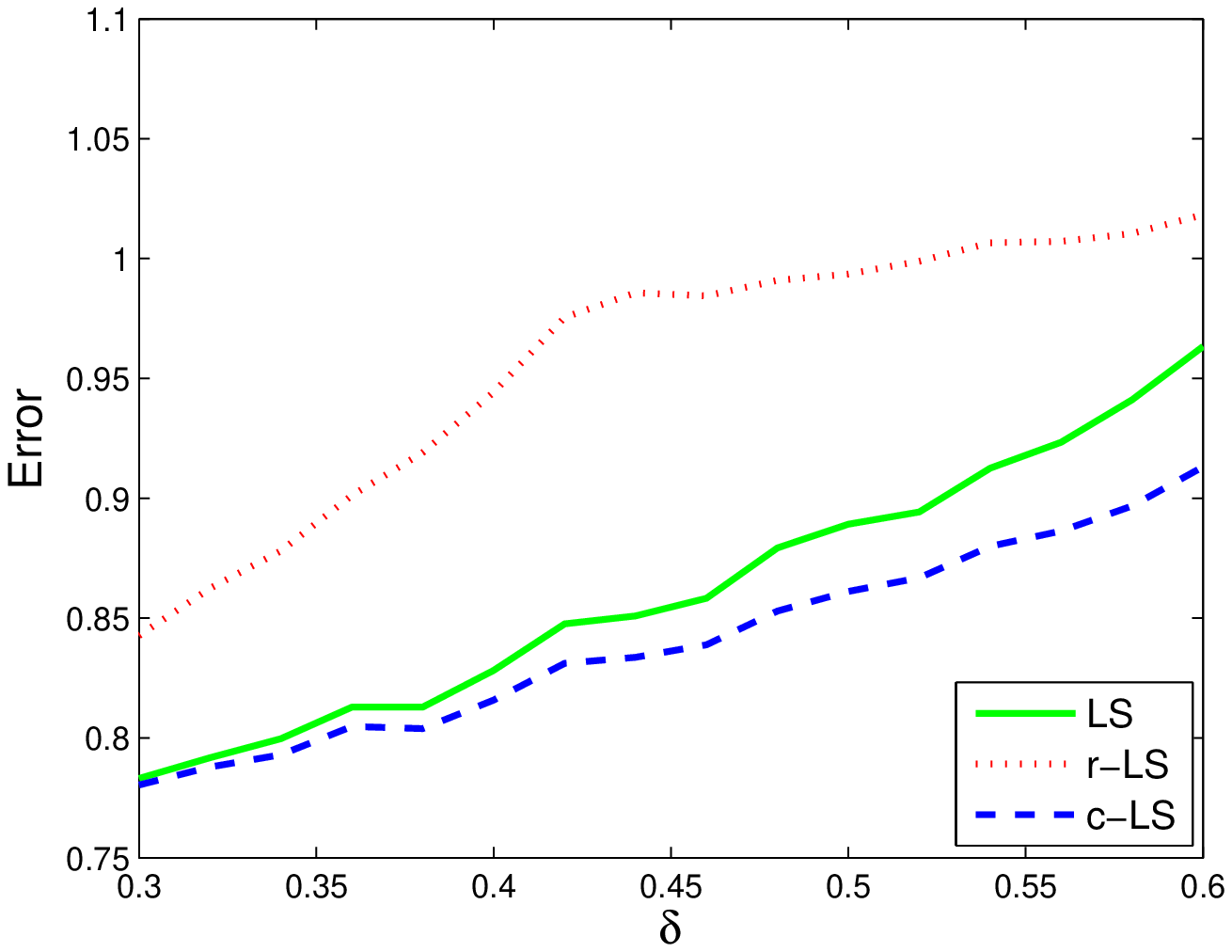}}
     \caption{\small Averaged errors for r-LS, c-LS and LS algorithms over 200 trials when $\rho \in [0.3, 06]$.}\vspace{-0.2cm}
     \label{fig:2}
  \end{center}
\end{minipage}
\hfill
\end{figure*}
Note that the LS algorithm yields the highest worst case error. Although the worst case error of the c-LS algorithm is
larger than the worst case error of the r-LS algorithm, the r-LS algorithm provides a superior
performance on the average with respect to both the r-LS and the LS algorithms.

For the second experiment, we randomly generate $200$ random perturbations $\dA$, $\dy$, where $\norm{\dA}\leq \rho_H$, $\norm{\dy}\leq \rho_Y$, $m=5$, $n=3$ for different perturbation bounds and compute the averaged error over $200$ trials for the c-LS, the LS, and the r-LS algorithms.
In Fig. 2, we present the averaged LS errors for the c-LS, the LS, and the c-LS algorithms where the perturbation bound varies, $\delta_Y=\delta_H = \delta \in [0.3, 0.6]$. We observe that the proposed c-LS algorithm has the best average LS error performance over different perturbation bounds compared to the LS and r-LS algorithms.
\begin{figure*}[t!]
\hfill
\begin{minipage}[t]{.46\textwidth}
  \begin{center}
  \vspace{-0.1cm}
  \epsfxsize=3.3in {\epsfbox{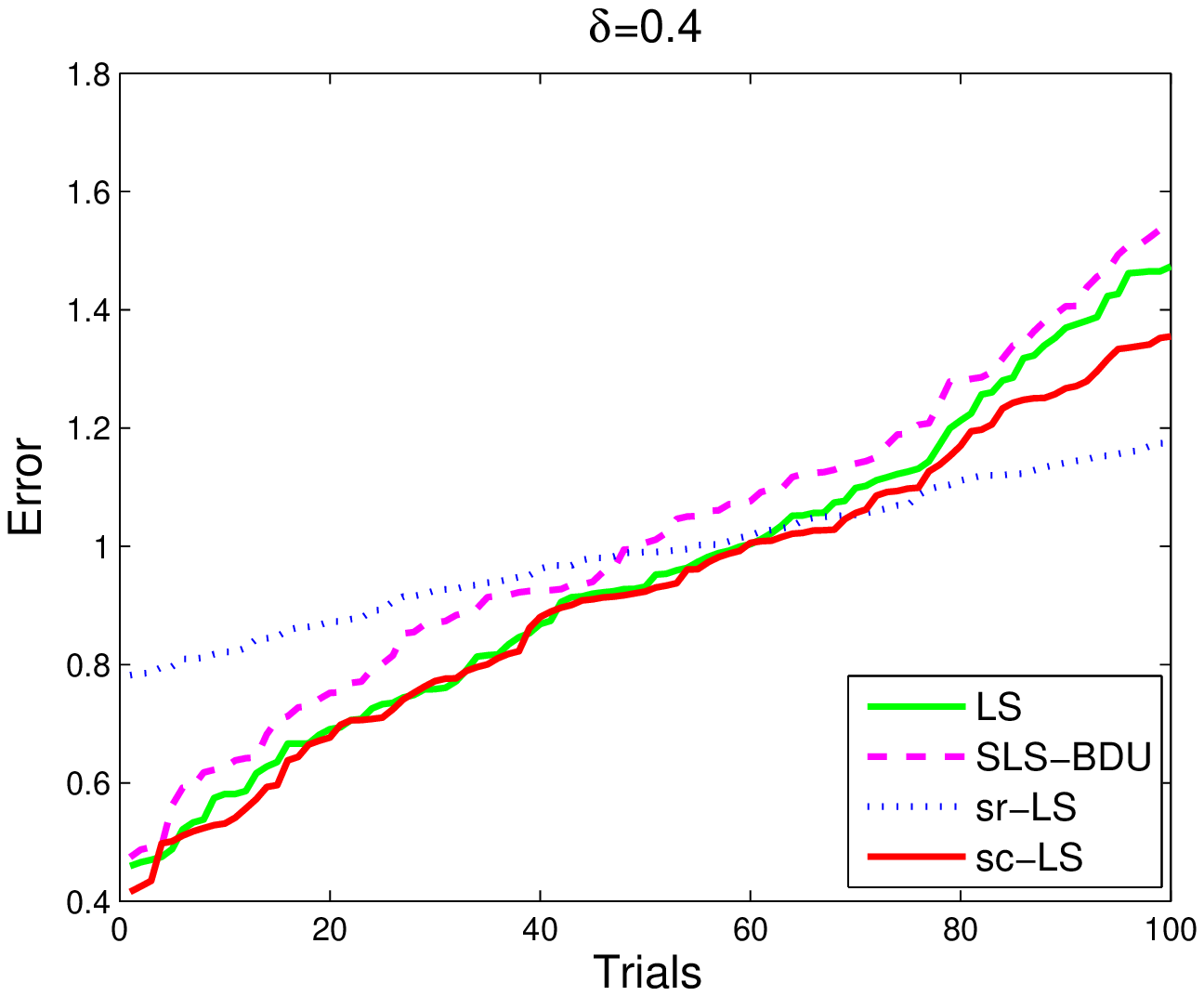}}
     \caption{\small Sorted errors for sr-LS, sc-LS, SLS-BDU, and LS algorithms over 100 trials when $\rho_H=\rho_Y=0.4$.}\vspace{-0.2cm}
   \label{fig:3}
   \end{center}
\end{minipage}
\hfill
\begin{minipage}[t]{.46\textwidth}
  \begin{center}
  \vspace{-0.1cm}
  \epsfxsize=3.3in {\epsfbox{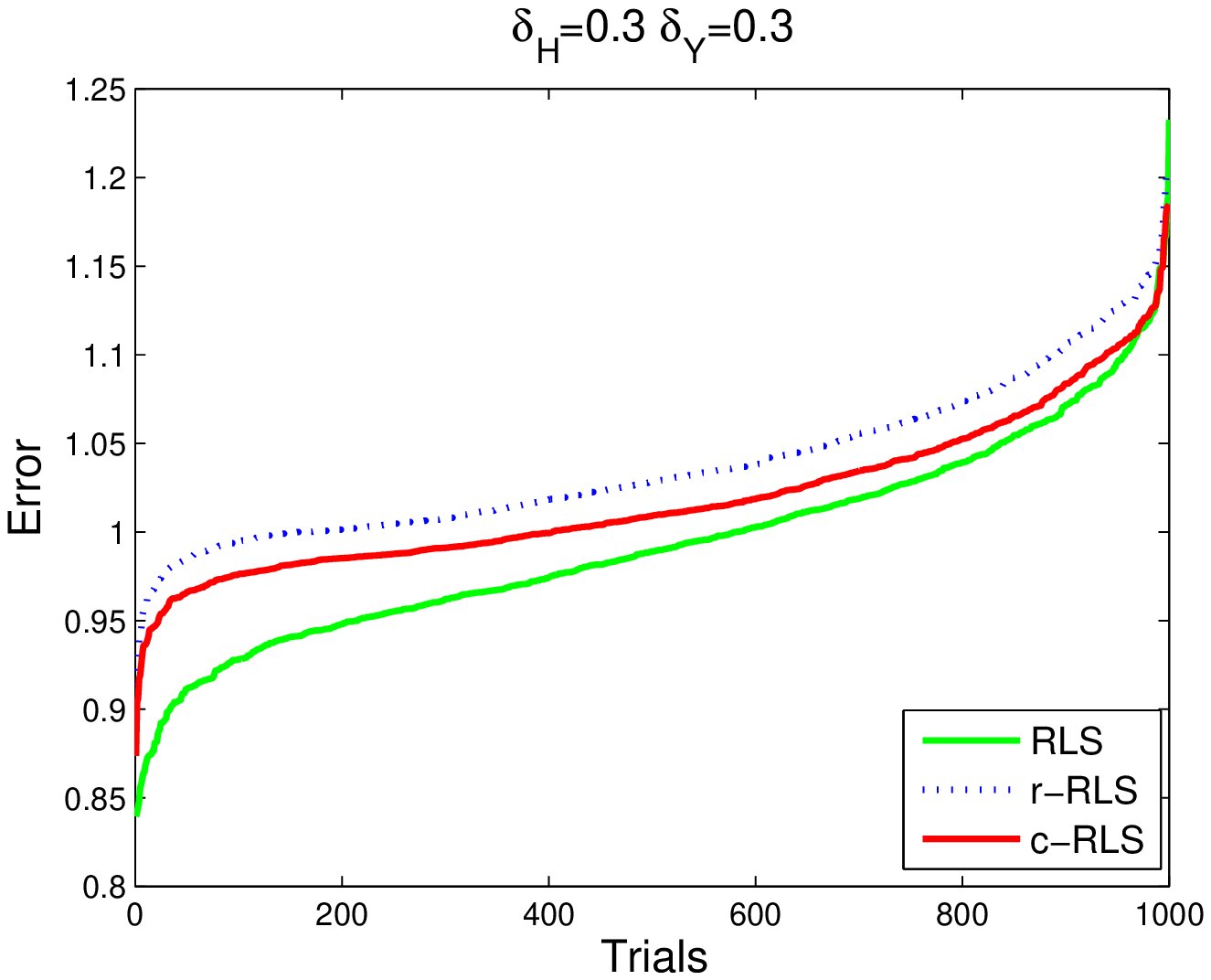}}
   \caption{\small Sorted errors for r-RLS, s-RLS and RLS algorithms over 1000 trials when $\rho_Y = \rho_H = 0.3$.}\vspace{-0.2cm}
    \label{fig:4}
  \end{center}
\end{minipage}
\hfill
\end{figure*}

In the next experiment, we examine a system identification problem
\cite{pilanci10}, which can be formulated as $\mU_0\vh=\vy_0$, where
$\mU=\mU_0+\mW$ is the observed noisy Toeplitz matrix and
$\vy=\vy_0+\vw$ is the observed noisy output vector. Here, the
convolution matrix $\mU$ (which is Toeplitz) constructed from $\vu$
which is selected as a random sequence of $\pm 1$'s. For a randomly
generated filter $\vh$ of length 3, we generate $100$ random
structured perturbations for $\mX_0$ and $\vy_0$, where
$\norm{\va}\leq 0.4 \norm{\mU_0}$, and plotted the sorted errors in
ascending order in Fig. 3. We observe that the largest errors are 1.53
for the structured least squares bounded data uncertainties, labeled
as ``{\it SLS-BDU}" and presented in \cite{pilanci10}, 1.47 for the LS
algorithm, 1.35 for the structured regret LS algorithm ``{\it sc-LS}"
from the Corollary, and 1.17 for the structured robust LS algorithm
``{\it sr-LS}". We observe that the sr-LS algorithm yields the largest
error on the average, however, it yields the smallest worst case error
among other algorithms as expected. In addition, we observe that the
sc-LS algorithm has a smaller worst case error compared to the LS and
the SLS-BDU algorithms, and it has the smallest average error compared
to the sr-LS, the SLS-BDU, and the LS algorithms, justifying its
competitiveness in these simulations. Finally, in Fig. 4, we provide
data errors sorted in ascending order for the algorithm in Theorem 2
as ``{\it c-RLS}", for the robust regularized LS algorithm in
\cite{sayed02} as ``{\it r-RLS}" and finally for the regularized LS
algorithm as ``{\it RLS}" \cite{Kailath:book}, where the experiment
setup is the same as in the first experiment except the number of
trials is 1000 and the perturbation bound is 0.3. The regularization
parameter is chosen as $\mu=0.1$. In these simulations, we observe
that the c-RLS algorithm trades off performance between the r-RLS and
the RLS algorithms. The RLS algorithm yields the largest error
compared to the r-RLS and c-RLS algorithms and
we observe that although c-RLS has an inferior performance compared to the RLS algorithm on the average it yields a superior performance than the r-RLS algorithm.
\section{Conclusion}
\label{sec:conc}
\vspace{-0.1cm} In this correspondence, we introduced a novel robust
approach to LS problems with bounded data uncertainties based on a
certain regret formulation.  We investigated the LS problems for both
unstructured and structured perturbations, and the robust regularized
LS problem for unstructured perturbations. In each case, the data
vectors that minimize the worst case regrets are found by solving
certain SDP problems. In our simulations, we observed that the
proposed algorithms provide a fair trade off between performance and
robustness, better than the best available alternatives in different
signal processing applications.  \appendix
\noindent{\bf Proof of Theorem 1}: By applying {\it S}-procedure \cite{boyd} to
 \eqref{prudefthm1}, and with some algebra one can show that \eqref{prudefthm1} is equivalent to solving the following SDP problem
\be
\min \lambda \mbox{ subject to}\nn\ee
\be
0\leq \left[ \begin{array}{cc}
 \lambda + \eta + \vc^T\vda + \vda^T\vc +\dy^T\vb + \vb^T\dy & \left(\tA\vx-\ty\right)^T\\
  \left(\tA\vx-\ty\right)  & \mI
 \end{array}  \right]
  \label{sdpesol2}.
\ee
Rearranging terms in \eqref{sdpesol2} results
\be
-\left[ \begin{array}{c}
 \vc^T \\
  \mX
 \end{array}  \right] \vda \left[ \begin{array}{cc}
 1 & \vec{0}
 \end{array}  \right]-\left[ \begin{array}{c}
 1 \\ \vec{0}
 \end{array}  \right] \vda^T \left[ \begin{array}{cc}
 \vc & \mX^T
 \end{array}  \right] \leq \left[ \begin{array}{cc}
 \lambda+ \eta + \dy^T\vb + \vb^T\dy& \left(\mA\vx-\ty\right)^T\\
  \left(\mA\vx-\ty\right)  & \mI
 \end{array}  \right]. \label{sdpesol3}
\ee
Here we used the identity $\dA\vx = \mX\vda$, where $\vda = \mathrm{vec}\left(\dA\right)$.
If we apply Proposition 2 of \cite{ElMe04} into \eqref{sdpesol3}, we obtain
\be
 -\left[ \begin{array}{c}
 \vb^T\\
  -\mI \\
  \vec{0}
 \end{array}  \right] \dy \left[ \begin{array}{ccc}
 1 & \vec{0} & \vec{0}
 \end{array}  \right]-\left[ \begin{array}{c}
 1 \\ \vec{0} \\ \vec{0}
 \end{array}  \right] \dy^T \left[ \begin{array}{ccc}
 \vb & -\mI & \vec{0}
 \end{array}  \right]\leq \left[ \begin{array}{ccc}
 \lambda+ \eta -\tau  & \left(\mA\vx-\vy\right)^T & \rho_H \vc^T \\
 \left(\mA\vx-\vy\right)  & \mI & \rho_H \mX \\
 \rho_H\vc & \rho_H\mX^T & \tau\mI
 \end{array}  \right]. \label{sdpesol5}
\ee
Applying Proposition 2 of \cite{ElMe04} again in \eqref{sdpesol5} yields  \eqref{sdpesol1}. $\square$
{\def\ninept{\def\baselinestretch{1.1}}
\vspace{-0.2cm}\bibliographystyle{IEEEbib}{\bibliography{bib_corresp}}
\end{document}}